\documentclass[%
 reprint,
 amsmath,amssymb,
 aps,
]{revtex4-2}

\usepackage{graphicx}
\usepackage{dcolumn}
\usepackage{bm}
\usepackage{subcaption}
\usepackage{hyperref}
\hypersetup{
    colorlinks=true,
    linkcolor=blue,
    filecolor=magenta,      
    urlcolor=cyan,
    citecolor=blue
    }

\begin{document}

\preprint{APS/123-QED}

\title{Revisit Many-body Interaction Heat Current and Thermal Conductivity Calculation in Moment Tensor Potential/LAMMPS Interface}

\author{Siu Ting Tai}
\author{Chen Wang}%
\author{Ruihuan Cheng}%
\author{Yue Chen}%
\thanks{yuechen@hku.hk}
\affiliation{
  Department of Mechanical Engineering, The University of Hong Kong, Pokfulam Road, Hong Kong SAR, China
}%

\date{October 30,2024}%

\begin{abstract}
The definition of heat current operator for systems for non-pairwise additive interactions and its impact on related lattice thermal conductivity ($\kappa_{L}$) via molecular dynamics simulation (MD) are ambiguous and controversial when migrating from conventional empirical potential models to machine learning potential (MLP) models. Empirical model descriptions are often limited to three- to four-body interaction while a sophisticated representation of the many-body physics could be resembled in MLPs. Herein, we study and compare the significance of many-body interaction to the heat current computation in one of the most popular MLP models, the Moment Tensor Potential (MTP) \cite{ivan2020}. Non-equilibrium MD simulations and equilibrium MD simulations among four different materials, PbTe, amorphous Sc\textsubscript{0.2}Sb\textsubscript{2}Te\textsubscript{3}, graphene, and BAs, were performed. We found inconsistency between the simulation thermostat and its implemented heat current operator in our non-equilibrium MD results which violate law of energy conservation and suggest a need for revision. We revisit the virial stress tensor expression within the calculator and identified the lack of a generalised many-body heat current description in it. We uncover the influence of the modified heat current formula that could alter the $\kappa_{L}$ results 29\% to 64\% using the equilibrium MD computational approach. Our work demonstrates the importance of a  many-body description during thermal analysis in MD simulations when MLPs are in concern. This work sheds light on a better understanding of the relationship between interatomic interaction and its heat transport mechanism. 
\end{abstract}

\maketitle%

\section{\label{sec:level1_intro}introduction\protect}

\par MLP has recently gained popularity in the application of computational materials engineering, especially for molecular dynamics (MD) simulations. Before the introduction of MLP, empiricial potential functions, such as Born-Mayer-Huggins (BMH) potential \cite{bornmayer1933}, Tersoff \cite{tersoff1986} potential, and Sillinger and Weber (SW) potential \cite{frankandweber1985} are implemented under a limited description to the interatomic interaction when used in conventional MD. They often fail to describe interactions of defects, surfaces, or metastable states. MLP alternatively provides a more generalized data-driven construction of the interatomic interaction model. Various established MLP models, for instance, neuroevolution potential (NEP) \cite{fan2021}, gaussian approximation potential (GAP) \cite{volker2021}, MTP \cite{ivan2020}, enable machine learning and training of the mathematics models coefficient over an expandable basis function set with desirable accuracy to encompass the complexity of many-body interaction within the local atomic environment neighborhood. MLP empowers more versatile applications such as simulations of the hydrogenation of amorphous silicon using GAP \cite{davis2022} and interfacial diffusion between Ge-Se alloy and Ti metal using MTP \cite{siddarth2022} that are not possible to accurately describe with conventional over-simplified potential due to complex physical interactions. MLPs are now a crucial utility in the simulation field for material reseachers to investigate novel materials and explore advanced applications.

\par A readily available software package, Machine Learning Interatomic Potential (MLIP), has been developed by Novikov et al. \cite{ivan2020} to address the need for a robust and computationally effective MLP. The MLIP features the MTP machine learning model with an interface to the software LAMMPS \cite{lammps2022}. Studies demonstrated the package as an effective interatomic potential in the research of thermal transport and phonon properties of different materials such as superionic conductor AgCrSe$_{2}$ \cite{wang2023}, wurtzite Boron Arsenide \cite{liu2021} and multiple 2D materials \cite{mortazavi2020}. Compared with other MLP models, MTP and its software package deliver an easy-to-use toolkit with better balance among performance, speed, and accuracy as compared with GAP, spectral Neighbor Analysis Potential (SNAP), and Neural Network Potential (NNP) \cite{zuo2020}. These advantages of the MLIP program introduce new opportunities to the material research community to predict lattice thermal conductivity ($\kappa_{L}$) and the thermal transport mechanism of a wide range of materials through MD simulation.

The study of thermal conductivity of materials is a crucial application of molecular simulation analysis in which MLPs are actively employed within the field. Experimental measurements often show challenges in determining heat transport properties of materials due to limitations in sample preparation and accuracy of empirical model. Theoretical computation results of $\kappa_{L}$ can be approximated through MD simulation approaches in which uncertainty are controlled. Three popular MD techniques, namely equilibrium MD (EMD), non-equilibrium MD (NEMD), and approach-to-equilibrium MD (AEMD) have been generally employed by researchers. The \textit{EMD} based on the Green-Kubo formula facilitates computation of $\kappa_{L}$ by relating heat current autocorrelation functions under the dissipation-fluctuation theorem as shown in 
\cite{zwanzig}.
\begin{equation}
\left.
\kappa_{L} =\frac{1}{3Vk_BT^2}\int_{0}^{\infty} \langle J(0)\cdot J(t) \rangle dt
\right.
\label{eqt:e1}
\end{equation}
where $\kappa_{L}$ , V , k\textsubscript{B} 
, T , and J represent the thermal conductivity coefficient, volume of simulation cells, Boltzmann's constant, temperature, and the heat current, respectively.

\par The \textit{NEMD} or direct method \cite{patrick2002} is conducted using a simulated, steady temperature gradient to study the heat current and $\kappa_{L}$ using Fourier’s law for heat conduction which is a direct analogous to an experimental measurement. A heat source and a heat sink sandwiched with a microcanonical \textit{NVE} ensemble domain are usually defined to provide steady heat current conditions. The third method, the approach-to-equilibrium technique \cite{lampin2013}, aims at simulating and capturing the out-of-equilibrium system responses of the temperature gradient when it returns to the equilibrium state. Two connecting domains, equilibrated at different temperatures, are modeled in a controlled simulation. The MD then allows energy flows between the two temperature blocks under a microcanonical \textit{NVE} simulation. By fitting the evolution to the temperature difference between the two blocks $\Delta T$ with the first exponential decay time $\tau$, the thermal conductivity of a material can be estimated.

\par Among the three methods, both EMD and NEMD are considered more popular in analyzing $\kappa_{L}$ of various material systems. The two techniques require the computation of the heat current of the simulation domain. Although there exist reports on the usage of MTP in determining $\kappa_{L}$ by MD simulation approaches \cite{huan2021}, there is a lack of focus on the computation of the heat current itself. Since both \textit{EMD} and direct approach require solving for the simulation model heat current element to obtain $\kappa_{L}$, it is crucial to understand how the MTP model evaluates the heat flux component of the system.

\par The heat current evaluation for MTP was implemented within the LAMMPS/MLIP interface but its deviation is ambiguous. The essence of the heat current calculation falls into Hardy’s expression of heat current in solid \cite{hardy1963}.

\begin{equation}
J= \sum\limits_{i}^{}\sum\limits_{j}^{}(r_i-r_j)(\frac{\partial E_i}{\partial r_j}\cdot v_j)
\label{eqt:e4}
\end{equation}
which $r_i$, $v_i$ and $E_i$ represent position, velocity and total energy of atom \textit{i}, respectively. This expression can be further simplified and was expressed as Eq. \ref{eqt:e5} in the LAMMPS package \cite{lammps2022,Boone2019}, where $e_i$ is the per-atom energy and $W_i$ is the atomic virial stress tensor.
\begin{equation}
J= \frac{1}{V}\sum\limits_{i}^{}e_i v_i-\sum\limits_{i}^{}W_i v_i
\label{eqt:e5}
\end{equation}
In the MLIP packages, the potential component of the heat current, presented as the second summation term on the right hand side in Eq. \ref{eqt:e5} was calculated as
\begin{equation}
J_{\text{pot}}= -\sum\limits_{i}^{}W_i v_i=\sum\limits_{i}^{}\sum\limits_{i\neq j}^{}r_{ij}(\frac{\partial U_i}{\partial r_{ij}}\cdot v_i)
\label{eqt:e6}
\end{equation}

\par Recently, a modification to the LAMMPS MD package brought new insight into how to handle and compute heat current in the simulation problem. Boone et al. \cite{Boone2019} and Surblys et al. \cite{surbly2019} suggested including the many-body interaction to the atomic stress formula, replacing pair-wise virial stress with a centroid virial stress tensor. The modification results in a higher heat current value in their example of butane, octane, and polystyrene which led to a highest reduction up to 25\% of $\kappa$ in butane. To transfer the idea of many-body heat current computation into MLP models, a generalized formula was developed \cite{fan2015} which is yet to be translated into other MLP systems. The potential contribution of the heat current can be written as Eq. \ref{eqt:e7}
\begin{equation}
J_{\text{pot}}= -\sum\limits_{i}^{}W_i v_i=\sum\limits_{i}^{}\sum\limits_{i\neq j}^{}r_{ij}(\frac{\partial U_j}{\partial r_{ji}}\cdot v_i)
\label{eqt:e7}
\end{equation}

\par It is worth noting that the expression to the atomic virial stress of atom \textit{i} is different with a subscript index of the atomic energy derivative term that swaps from \textit{i} to \textit{j} when comparing equation \ref{eqt:e6} with \ref{eqt:e7}. The derivation of the many-body heat current formula suggested that the original heat current expressed by the MLIP was a pair-wise interaction expression that assumes a premise of $U_i=U_j$ that energy between a pair of atoms only depends on their distance $r_{ij}$ as discussed in Ref. \cite{fan2015}. Although MTP potential provides a mathematical model that considers the many-body interactions between atomic local environment when evaluating the energy and forces of the system, the MLIP software construction shows a lack of ability to express the heat current under a many-body context. We intend to demonstrate the significance of employing the generalized many-body heat current formula with examples of various materials in this paper. 

\section{\label{sec:level1_comp}Computation Details for MD Simulations\protect}
\begin{figure*}%
    \includegraphics[scale=0.15]{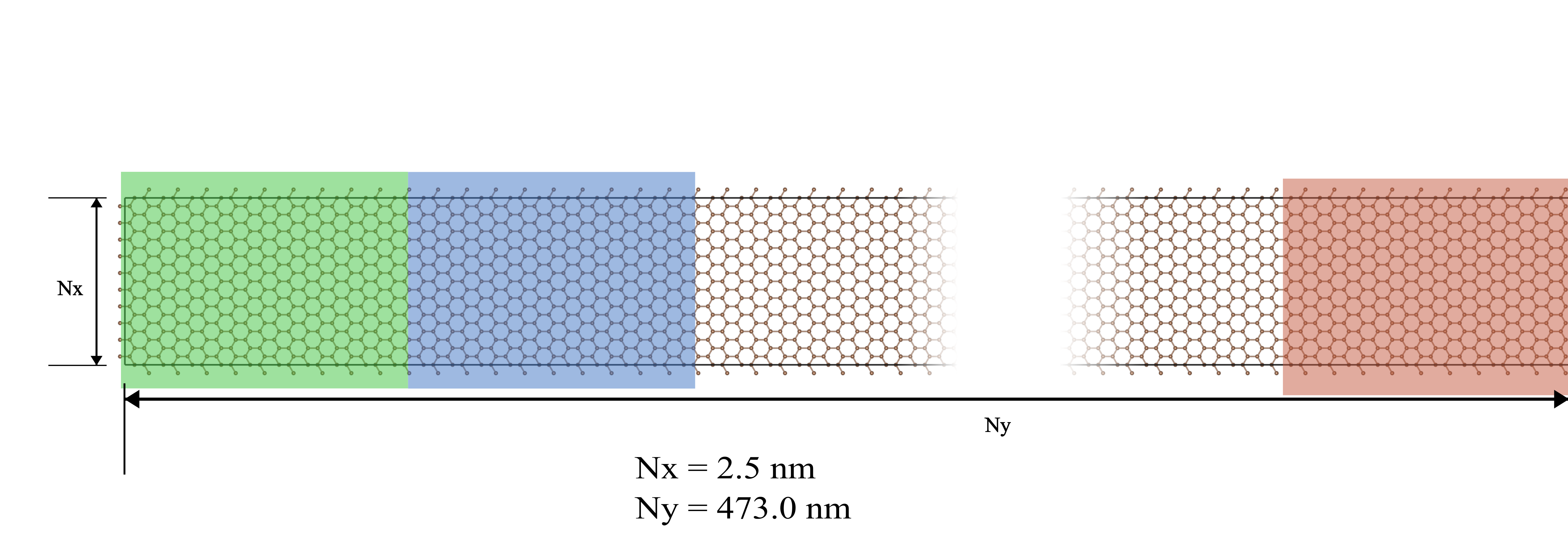} %
    \caption{A simplified single-layered graphene model for \textit{NEMD} simulation. Green, blue and red sections represent the fixed end, the heat sink and the heat source, respectively.}%
    \label{fig:im}%
\end{figure*}

\par With our aim to illustrate the needs and effect of modifying the heat current description to express the non-pairwise contribution in MTP, we perform four sets of MD simulations over different materials, namely PbTe, amorphous Sc\textsubscript{0.2}Sb\textsubscript{2}Te\textsubscript{3}, graphene, and BAs. These examples were selected to cover a range of order of magnitude from $10^{-1}$ to $10^3$ W/mK in $\kappa_{L}$ and a variety of spatial complexity and symmetry order. The four chosen materials are able to generally validate the potential impact of the generalized many-body heat current formula to the computed heat current and its influence on $\kappa_{L}$. Both \textit{NEMD} and \textit{EMD} were performed for each of the mentioned material examples to demonstrate the significance of the heat current formulation modification in this work.

\par MTP interatomic potentials for the four respective materials are prepared by a passive MTP training utilizing the MLIP packages. Training sets for PbTe, amorphous Sc\textsubscript{0.2}Sb\textsubscript{2}Te\textsubscript{3}, and graphene are extracted from established works of Cheng et al. \cite{rhchenf2023}, Wang et al. \cite{wang2024}, and Rowe et al. \cite{rowe2020}, respectively. The training set for BAs was prepared independently in a similar manner described in Mortazavi’s work \cite{mortazavi2021b} through sampling atomic structure configurations in ab-initio MD (AIMD) simulation trajectories of a $5\times5\times5$ supercell of BAs using the Vienna Ab-initio Simulation Package (VASP) \cite{kresse1996}. Forces and energies of 408 configurations sampled from 200-900 K were evaluated under the density functional theory (DFT) framework using the VASP package. The exchange-correlation functional was approximated using the projector-augmented wave (PAW) method \cite{paw1994} incorporated with the Perdew-Burke-Ernzerhof (PBE) generalized gradient approximation (GGA). A $2\times2\times2$ $\Gamma$-centered \textit{k}-point mesh with a kinetic energy cutoff of 600 eV was employed. The energy convergence threshold was set to be $10^{-7}$ eV for all the single point self-consistent energy calculations. 

\par The \textit{NEMD} simulation was employed to calculate the room temperature heat current for PbTe, amorphous Sc\textsubscript{0.2}Sb\textsubscript{2}Te\textsubscript{3}, graphene, and BAs using the MTP potentials trained respectively. We adopted a similar heat current validation approaches in Ref. \cite{dong2024} to uncover the necessity of expressing the heat current through the many-body heat current formula when incorporating the MTP potential. A steady temperature gradient centered at 300 K along the \textit{y}-direction of the periodic orthogonal models was employed with a fixed end to insulate heat flow between the heat sink and source. Fig. \ref{fig:im} shows an example of the schematic diagram of a single-layered graphene model of 44400 atoms. The model spans a length of around 473 nm, a width of 2.5 nm, and a 3 nm vacuum space in the out-of-plane direction, which we found to be a decent balance between the simulation dimension with at least 300 nm in the characteristic heat flow direction and a computational expense of around 10000 cpu hours for each simulation.

\par The other two bulk material models of PbTe and amorphous Sc\textsubscript{0.2}Sb\textsubscript{2}Te\textsubscript{3} were structured in similar configurations with reduced cell sizes since their expected thermal conductivities are significantly lower with a shorter phonon mean free path when compared to graphene. The lengths of the simulation cells were chosen to be 32 nm, and 45.7 nm, respectively. The amorphous Sc\textsubscript{0.2}Sb\textsubscript{2}Te\textsubscript{3} model was built to contain a much higher number of atoms of 18000 due to a larger unit cell for representing a disordered structure. The model of BAs was scaled to have length of 486 nm which is similar with the graphene structure as their room temperature phonon mean free path are comparable at the range of 2 to 10 $\mu$m \cite{broido2013,fugallo2014}. 

\par All four sets of \textit{NEMD} simulations were carried out in two stages. A time step of 1 fs was chosen for all four models. The simulation domain, except the fixed end section, was first equilibrated at the center temperature of 300 K for 200 ps under an \textit{NVT} canonical ensemble. The system was then run for another 2 ns under an \textit{NVE} ensemble with a steady and uniform temperature gradient applied to the heat source and heat sink through a Langevin thermostat at 350 K and 250 K, respectively. We considered the first 0.5 ns of the run to stabilize the system under such a temperature gradient and recorded the heat current accounted for by the Langevin heat baths and the calculated heat current through the MLIP interface for the remaining 1.5 ns of the run.

\par Another angle to visualize the impact of using the many-body heat current formula in the MTP potential is the change in $\kappa_{L}$ computed by the \textit{EMD} methods. This technique involves intensive evaluation of heat current using the MLIP interfaces and brings insight into its legitimacy. The numbers of atoms in each model were selected to be 1728, 11520, 1800, and 1728 for PbTe, amorphous Sc\textsubscript{0.2}Sb\textsubscript{2}Te\textsubscript{3}, graphene, and BAs, respectively. Five independent simulations, each containing 10 complete auto-correlation windows, were performed on each material. The first four correlation windows were discarded and 30 sampled heat current auto-correlation functions were completed for all four examples. The correlation times were chosen to be 40 ps and 80 ps for PbTe and amorphous Sc\textsubscript{0.2}Sb\textsubscript{2}Te\textsubscript{3}, respectively. Graphene and BAs have higher $\kappa_{L}$ and use a correlation time of 1000 ps. The simulation time step was set to be 1 fs. All simulations were first equilibrated at 300 K under an \textit{NVT} ensemble and later ran a length covering 10 correlation time windows under the \textit{NVE} ensemble. The heat current auto-correlation function was calculated in the latter half of the MD simulations as implemented within the LAMMPS package and we computed the $\kappa_{L}$ of each material subjected to both the original and the modified heat current formula using Eq. \ref{eqt:e1}.

\section{\label{sec:level1_result}Results and Discussions\protect}

\subsection{\label{sec:level2_nemd}\textit{NEMD} Simulation for Direct Heat Current Evaluation}

\par We first consider the cubic PbTe crystal. The \textit{NEMD} simulation result in Fig. \ref{fig:sub1}(a) illustrates the overall cumulative heat passed through the heat source and sink as orange dash line, which was evaluated within the LAMMPS implementation of the ensemble thermostat. On the other hand, the cumulative heat evaluated by MTP across the simulation domain, as drawn in blue line, deviates from the overall \textit{NEMD} heat current balance. The result shows strong evidence that the original implementation of virial stress heat current formula that involves only pairwise interaction fails to represent the accurate heat current component in the MD simulation, undermining the validity of using the MLIP software in computing $\kappa_{L}$ using the MD approaches. Comparing Fig. \ref{fig:sub1}(a) and (b), the introduction of the many-body heat current formula significantly improves energy balance between heat current within the simulation thermostat system and the intrinsic heat current of the sandwiched region evaluated by MTP. These simulations suggest that the original method underestimates the actual heat current by around 50\% and could impact the result of $\kappa_{L}$ calculation by direct method MD simulations.

\def \mywidth {7}
\def \myheight {7}
\begin{figure}%
    \centering
    \subfloat[Original]{{\includegraphics[width=\mywidth cm,height=\myheight cm]{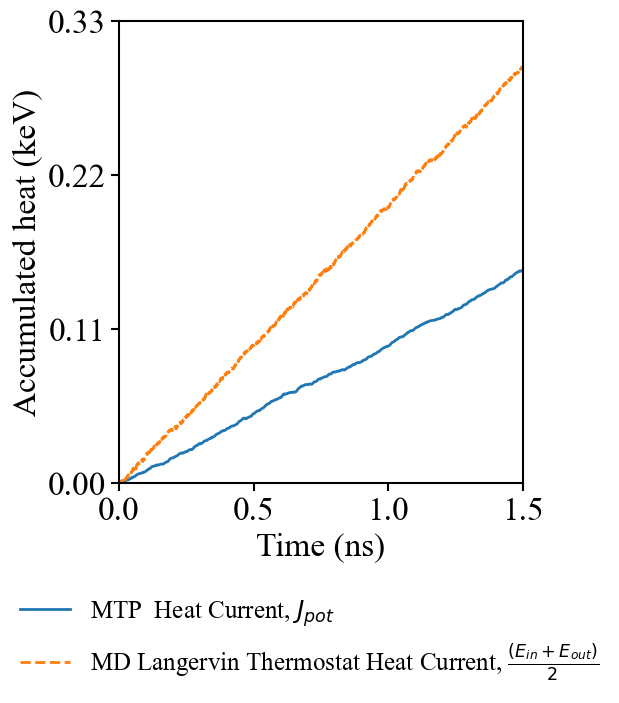} }}\hspace{0.0cm}%
    \subfloat[Modified]{{\includegraphics[width=\mywidth cm,height=\myheight cm]{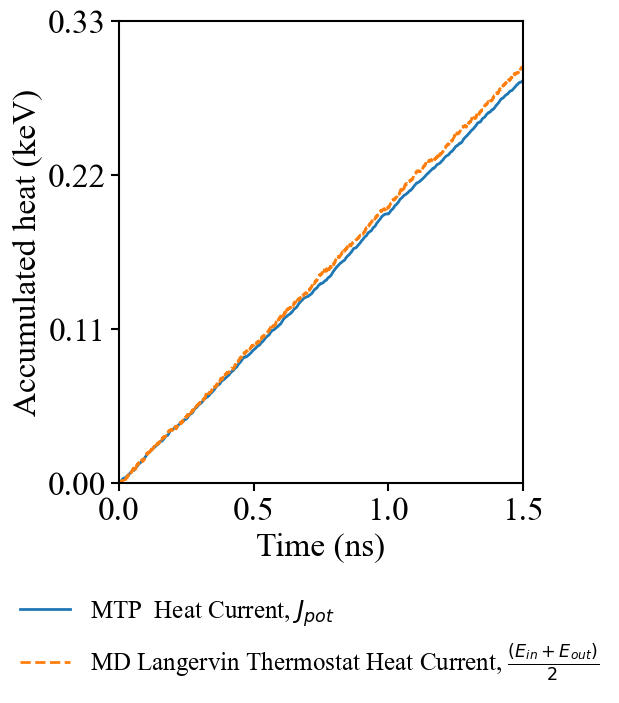} }}%
    \caption{Accumulative heat of \textit{NEMD} simulations of PbTe using (a) original MLIP interface and (b) modified MLIP interface with many-body heat current formula correction.} 
    \label{fig:sub1}%
\end{figure}
\begin{figure}%
    \centering
    \subfloat[Original]{{\includegraphics[width=\mywidth cm,height=\myheight cm]{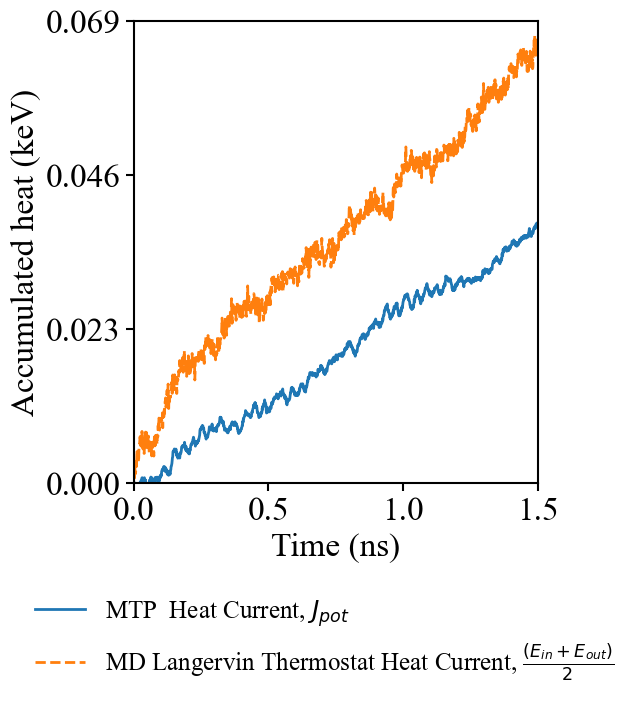} }}\hspace{0.0cm}%
    \subfloat[Modified]{{\includegraphics[width=\mywidth cm,height=\myheight cm]{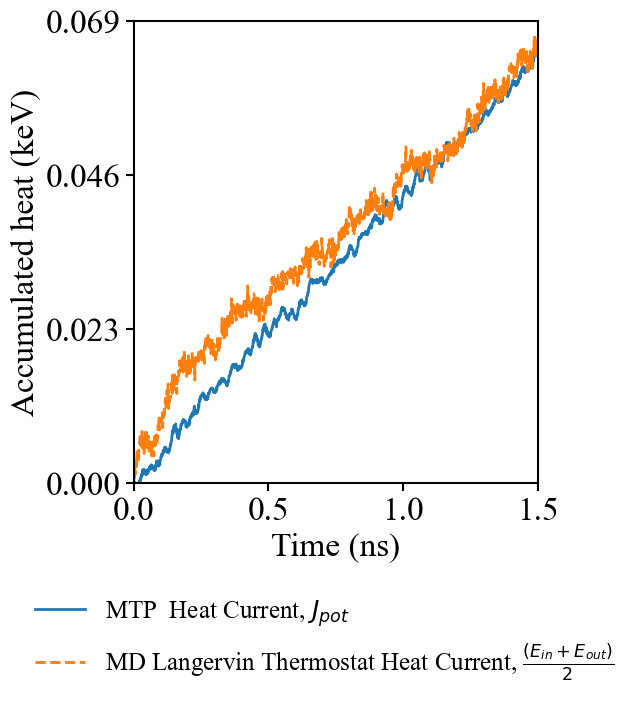} }}%
    \caption{Accumulative heat of \textit{NEMD} simulations of amorphous Sc\textsubscript{0.2}Sb\textsubscript{2}Te\textsubscript{3} using (a) original MLIP interface and (b) modified MLIP interface with many-body heat current formula correction.}
    \label{fig:sub2}%
\end{figure}
\begin{figure}%
    \centering
    \subfloat[Original]{{\includegraphics[width=\mywidth cm,height=\myheight cm]{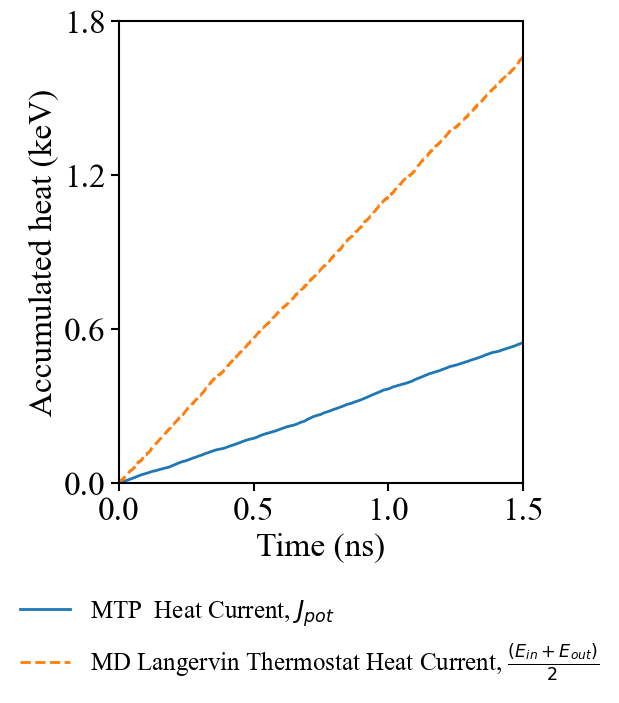} }}\hspace{0.0cm}%
    \subfloat[Modified]{{\includegraphics[width=\mywidth cm,height=\myheight cm]{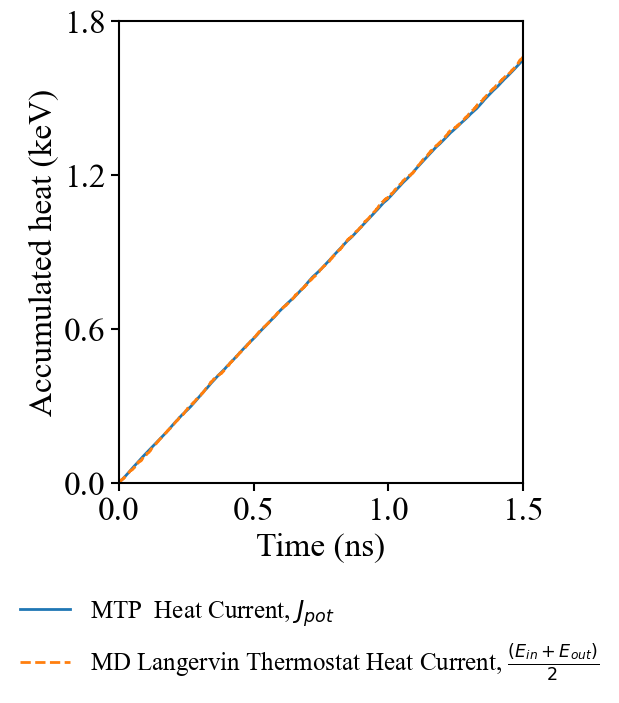} }}%
    \caption{Accumulative heat of \textit{NEMD} simulations of graphene using (a) original MLIP interface and (b) modified MLIP interface with many-body heat current formula correction.}
    \label{fig:sub3}%
\end{figure}
\begin{figure}%
    \centering
    \subfloat[Original]{{\includegraphics[width=\mywidth cm,height=\myheight cm]{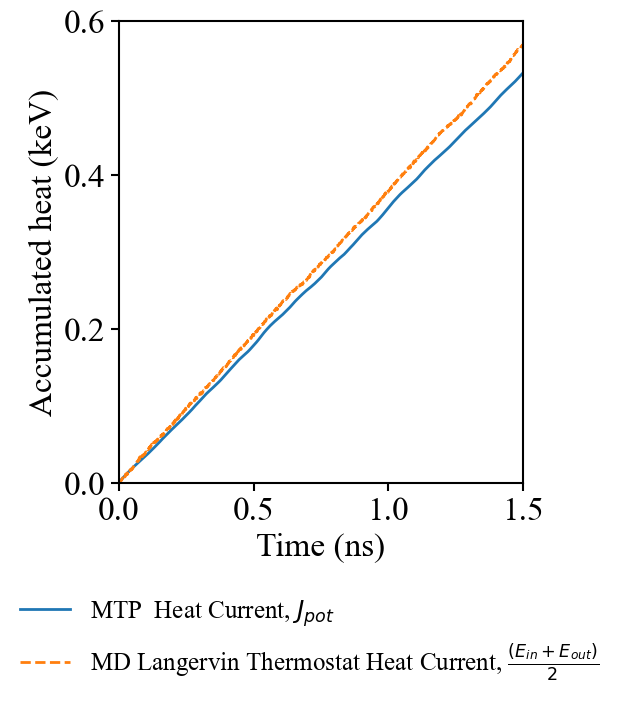} }}\hspace{0.0cm}%
    \subfloat[Modified]{{\includegraphics[width=\mywidth cm,height=\myheight cm]{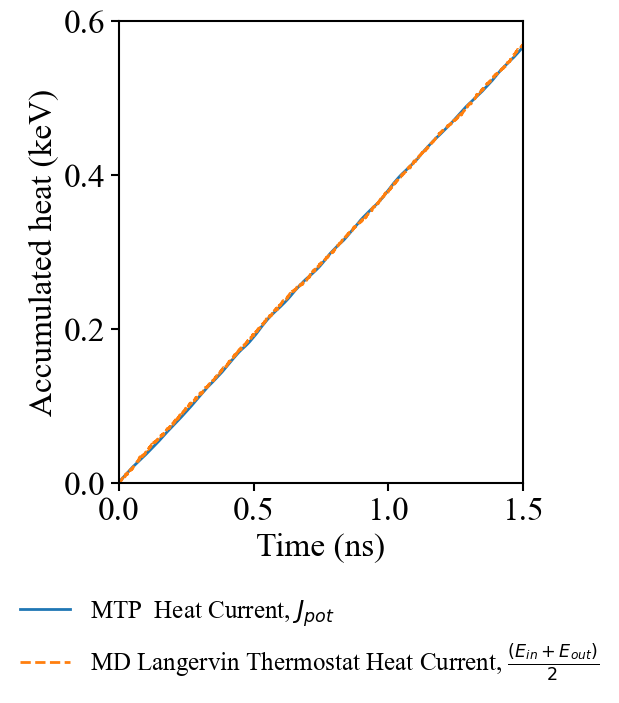} }}%
    \caption{Accumulative heat of \textit{NEMD} simulations of BAs using (a) original MLIP interface and (b) modified MLIP interface with many-body heat current formula correction.}
    \label{fig:sub4}%
\end{figure}

\par A more complex material with a disordered structure of amorphous Sc\textsubscript{0.2}Sb\textsubscript{2}Te\textsubscript{3} shows a similar trend to PbTe, suggesting that the issue of the MTP original heat current representation is general regardless of the structure complexity of the subject. This ternary phase change material was constructed by melt and quench of a large unit cell consisting of 180 atoms with a lattice parameter of length up to 18 \r{A}, assuring the randomness of atomic configurations well within the MTP interaction cutoff radius. A similar comparison of the cumulative heat current from the MD simulation heat bath and the evaluated heat current within the simulated domain is shown in Fig. \ref{fig:sub2}. The original heat current implementation underestimates the heat current by around 40\% while the many-body heat current correction shows better agreement between the simulation heat bath and the heat current computed by the MTP potential. This result also indicates a significant improvement in the evaluation of heat current using the many-body formula under a disordered and locally more sophisticated atomic neighborhood environment.

\par Then, we extend the research to the effect of the many-body heat current correction on the 2D material graphene. The cumulative heat current of the non-equilibrium simulations suggests the existence of a similar discrepancy between the average heat current on the ensemble thermostats and the heat current evaluated by the virial stress heat current formulation of the LAMMPS/MLIP interface package across the simulation domain. The original formula leads to the biggest difference among four testing system, 75\% lower than the MD simulation thermostats. The modified many-body heat current formula shows a significant improvement between the heat baths' overall heat current and the MTP potential. This result agrees with the recent study of graphene by Dong et al. \cite{dong2024}, where MTP potential model was compared with other neural network based models which had adapted to the many-body heat current interation description. Fig. \ref{fig:sub3} provides key evidence to the incapability of the original implementation in the MLIP package in determining the heat current in a 2D system.

\par Finally, we find in the non-equilibrium simulation that BAs behaves differently from the previous three materials. The cumulative heat current of the original implementation in Fig. \ref{fig:sub4} reveals a slight underestimation of heat current computed by the original heat current expression. The modified many-body heat current results exhibit better agreement among the overall heat current evaluated by the MD simulation thermostat and the MTP computed heat current, while the original pair-wise heat current reveals a small deviation in the 1.5 ns time frame. This implies the possibility that the many-body interaction component in the potential component to the heat current is less significant in the case of BAs.

\subsection{\label{sec:level2_emd}\textit{EMD} Simulation for Thermal Conductivity}

\par We then compare the effect on $\kappa_{L}$ computed based on the Green-Kubo method using the \textit{EMD} approach. Under the same simulation cell size and constraint, PbTe shows a 64\% increase of $\kappa_{L}$ from 0.97 to 1.59 W/mK when comparing the original heat current formula and the modified many-body heat current formula. Fig. \ref{fig:sub5} shows the thermal conductivity against the correlation time of PbTe during \textit{EMD}. The modified formula shows better agreement with experimental result \cite{pei2012} of 1.52 W/mK, suggesting that the migration to many-body heat current implementation is a more complete description for the MTP potential.
\def \mywidth {7.5}
\def \myheight {5}
\begin{figure}%
    \centering
    \subfloat[Original]{{\includegraphics[width=\mywidth cm,height=\myheight cm]{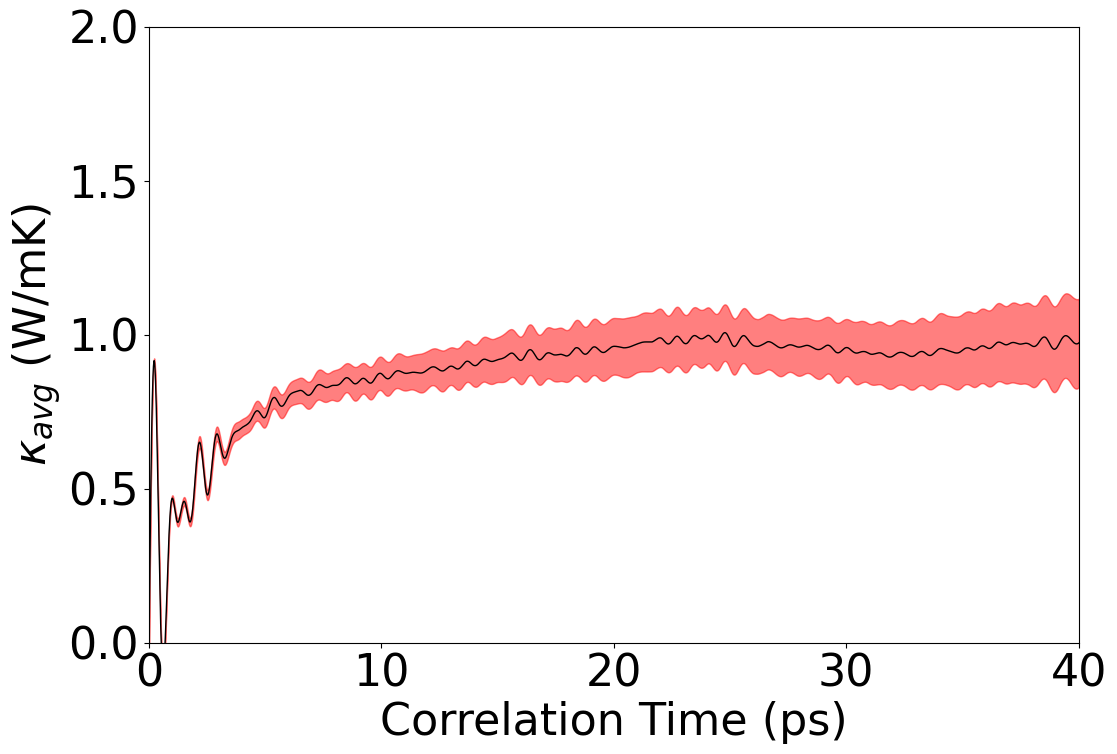} }}%
    \qquad%
    \subfloat[Modified]{{\includegraphics[width=\mywidth cm,height=\myheight cm]{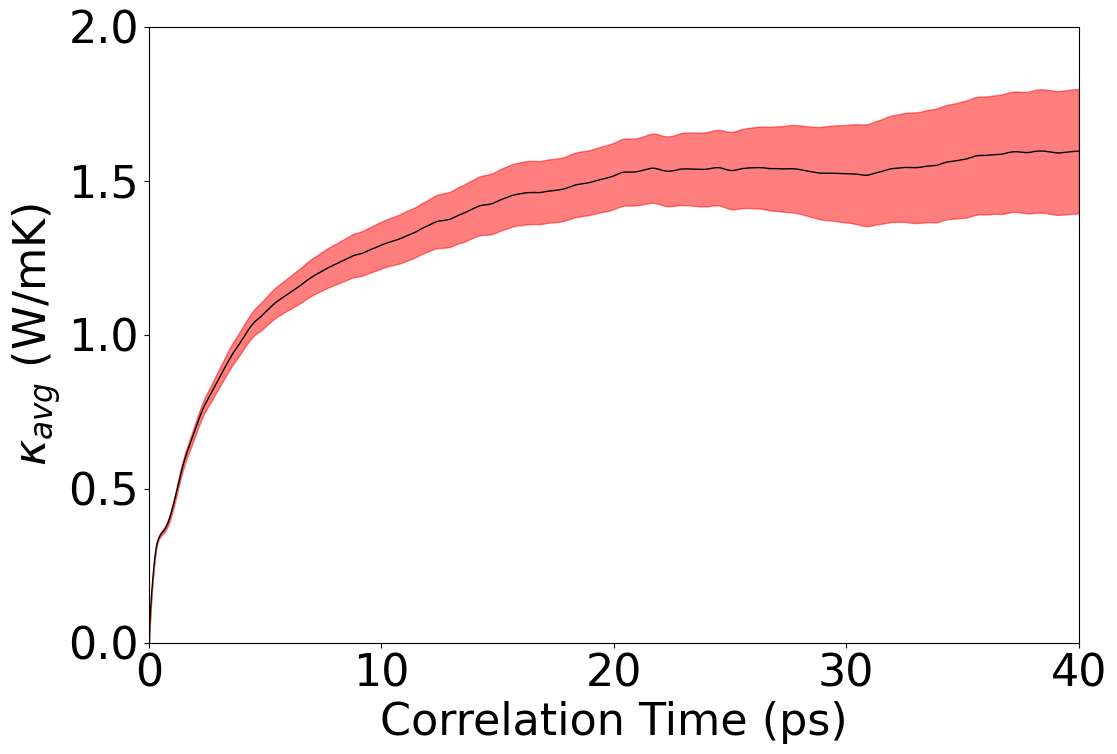} }}%
    \caption{Averaged $\kappa_{L}$ of PbTe at 300K from 30 independent heat auto-correlation functions against correlation time using (a) original MLIP interface and (b) modified MLIP interface with many-body heat current formula correction. Shaded red area represents upper and lower bounds within one standard deviation.}%
    \label{fig:sub5}%
\end{figure}
\begin{figure}%
    \centering
    \subfloat[Original]{{\includegraphics[width=\mywidth cm,height=\myheight cm]{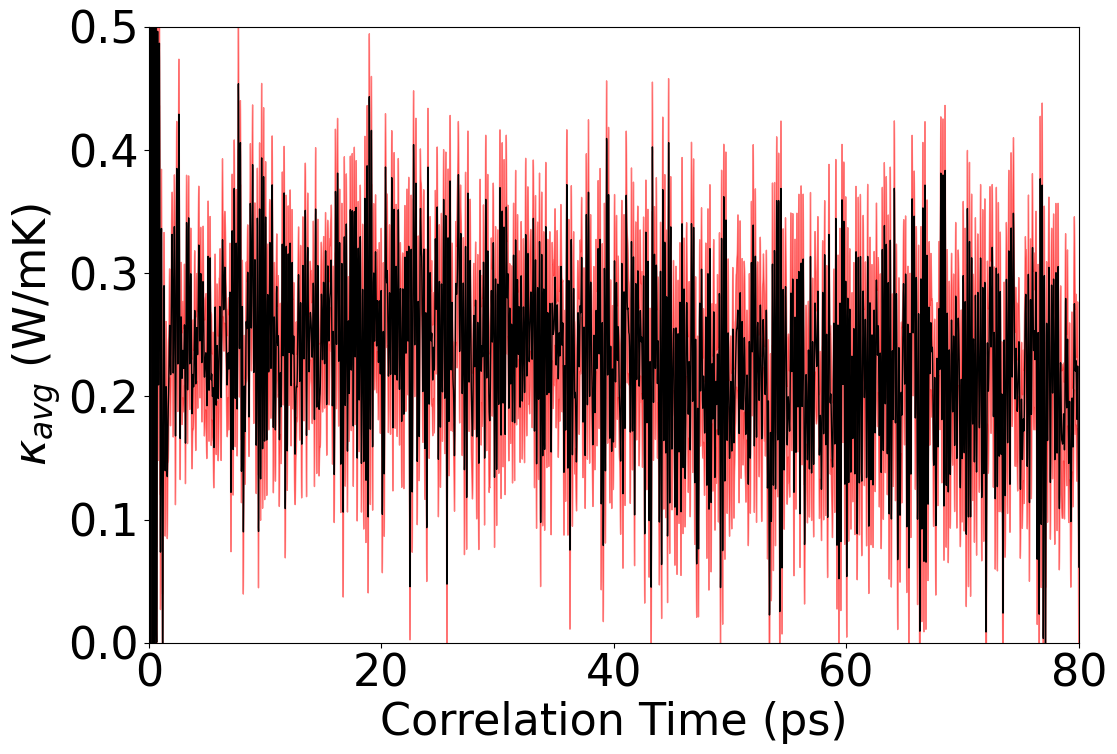} }}%
    \qquad
    \subfloat[Modified]{{\includegraphics[width=\mywidth cm,height=\myheight cm]{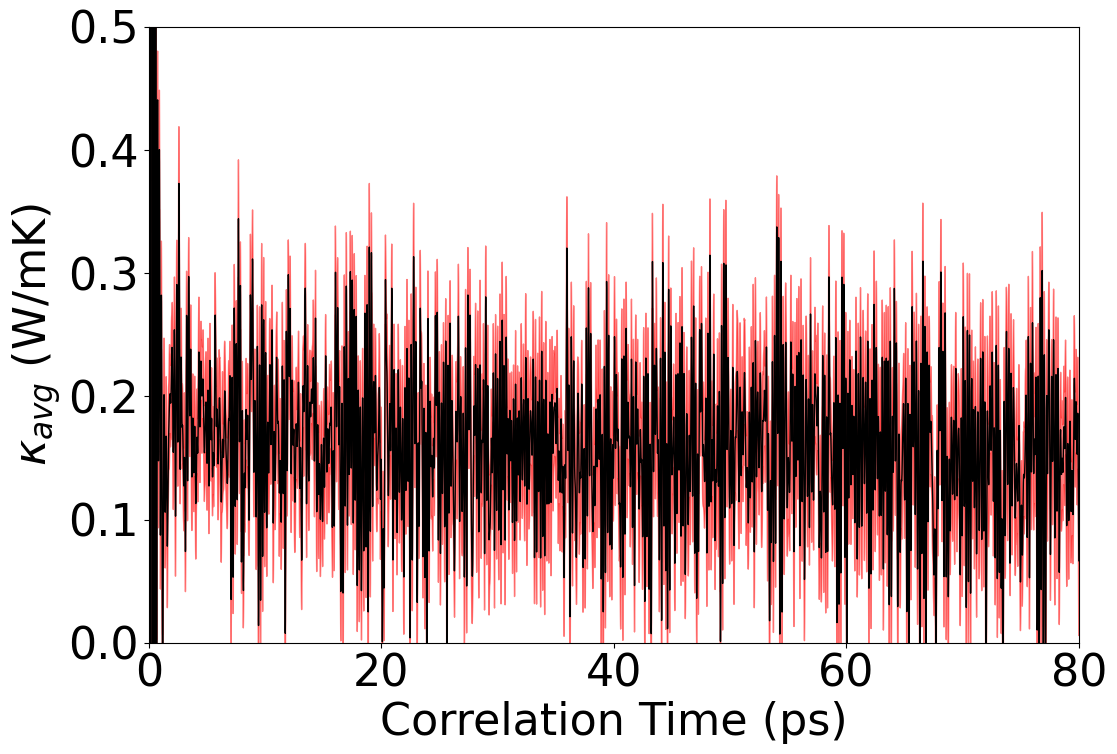} }}%
    \caption{Averaged $\kappa_{L}$ of amorphous Sc\textsubscript{0.2}Sb\textsubscript{2}Te\textsubscript{3} at 300K from 30 independent heat auto-correlation functions against correlation time using (a) original MLIP interface and (b) modified MLIP interface with many-body heat current formula correction. Shaded red area represents upper and lower bounds within one standard deviation.}%
    \label{fig:sub6}%
\end{figure}
\begin{figure}%
    \centering
    \subfloat[Original]{{\includegraphics[width=\mywidth cm,height=\myheight cm]{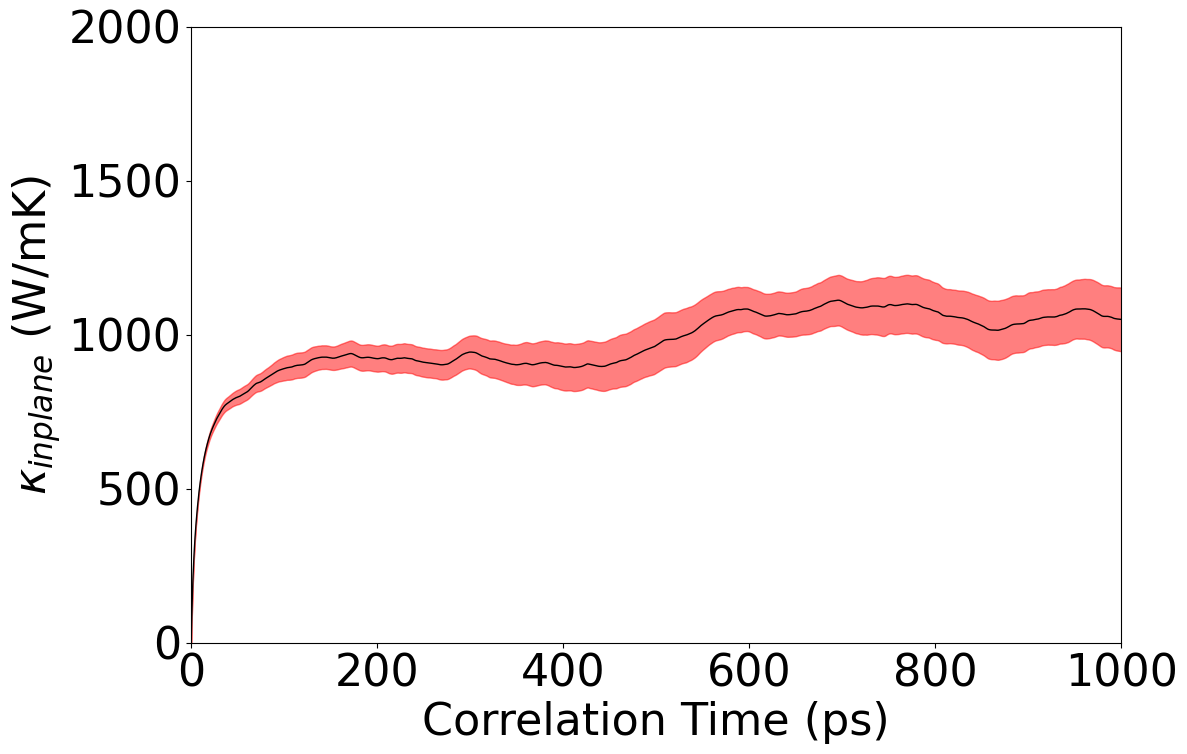} }}%
    \qquad
    \subfloat[Modified]{{\includegraphics[width=\mywidth cm,height=\myheight cm]{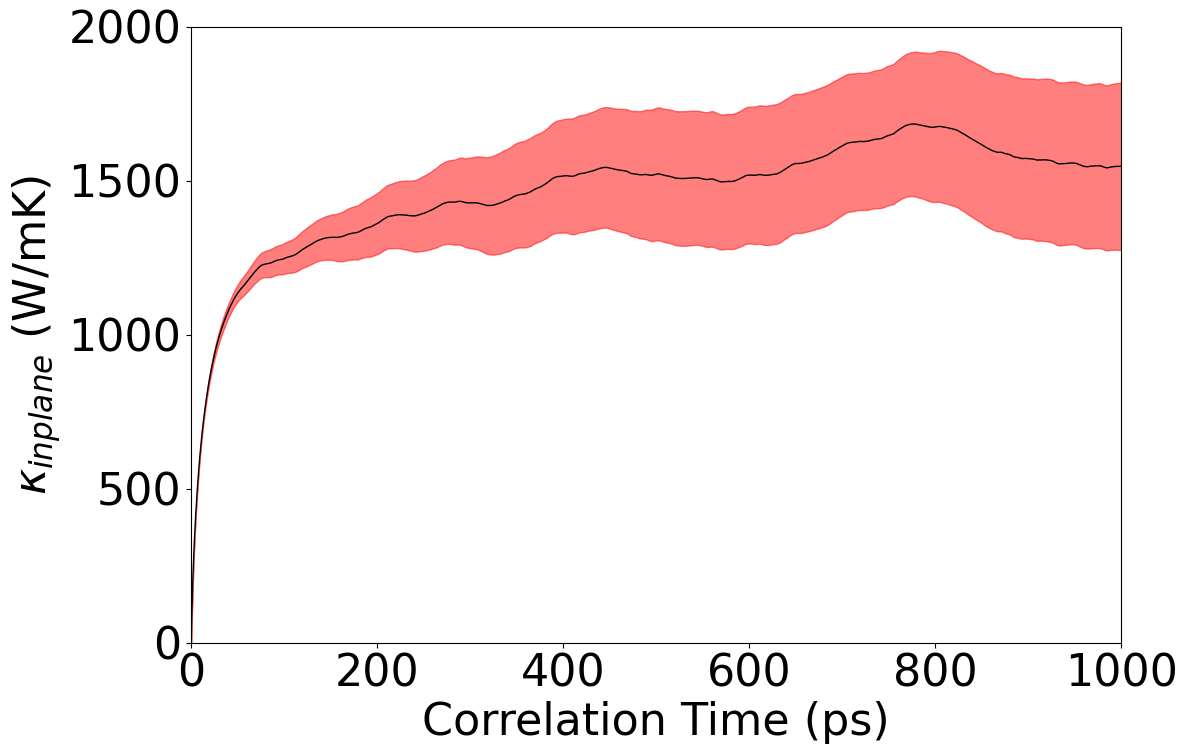} }}%
    \caption{Averaged $\kappa_{L}$ of graphene at 300K from 30 independent heat auto-correlation functions against correlation time using (a) original MLIP interface and (b) modified MLIP interface with many-body heat current formula correction. Shaded red area represents upper and lower bounds within one standard deviation.}%
    \label{fig:sub7}%
\end{figure}
\begin{figure}%
    \centering
    \subfloat[Original]{{\includegraphics[width=\mywidth cm,height=\myheight cm]{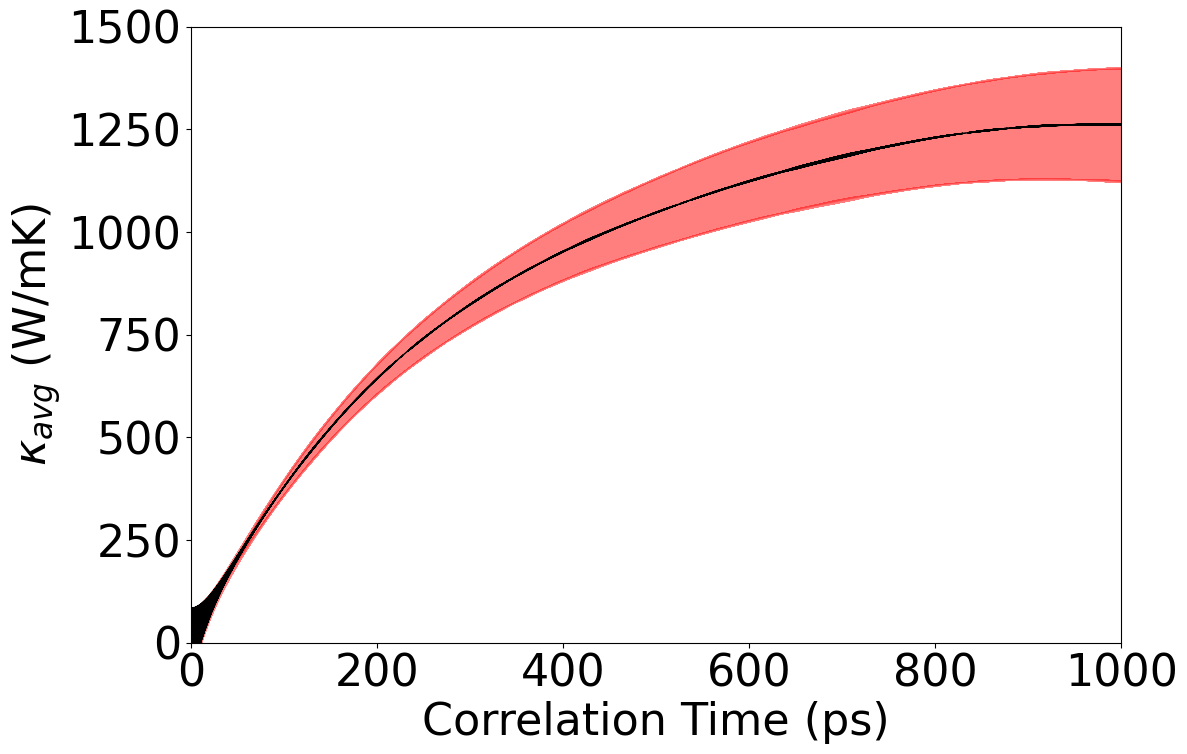} }}%
    \qquad
    \subfloat[Modified]{{\includegraphics[width=\mywidth cm,height=\myheight cm]{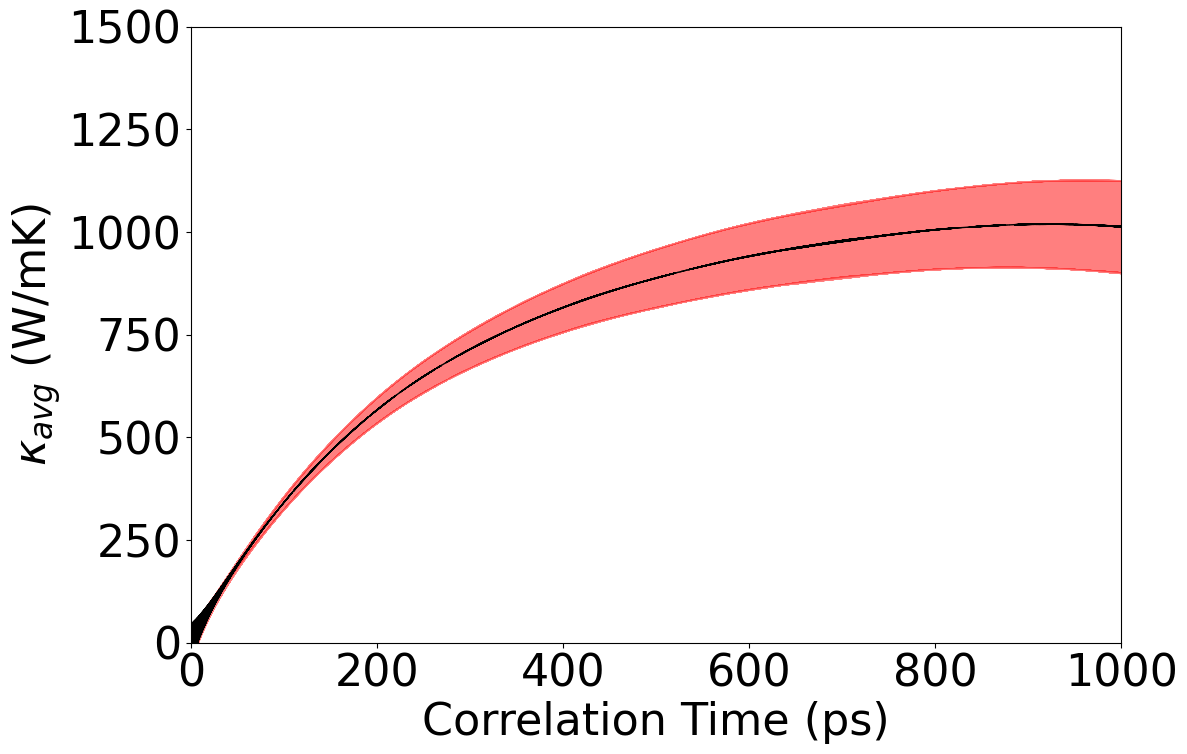} }}%
    \caption{Averaged $\kappa_{L}$ of BAs at 300K from 30 independent heat auto-correlation functions against correlation time using (a) original MLIP interface and (b) modified MLIP interface with many-body heat current formula correction. Shaded red area represents upper and lower bounds within one standard deviation.}%
    \label{fig:sub8}%
\end{figure}
\par $\kappa_{L}$ of amorphous Sc\textsubscript{0.2}Sb\textsubscript{2}Te\textsubscript{3} was then evaluated which revealed a decrease in $\kappa_{L}$ from 0.21 to 0.15 W/mK. The result shows better agreement with Ref. \cite{wang2024}, in which $\kappa_{L}$ was computed with Allen-Feldman and sinusoidal approach-to-equilibrium MD (SAEMD) approaches. Fig. \ref{fig:sub6} reveals a noisy yet distinguishable change between the modifications. Regarding the previous example of PbTe, one may anticipate a similar increase in $\kappa_{L}$ with a more complete heat current contribution from the many-body interaction component. However, in this case of amorphous Sc\textsubscript{0.2}Sb\textsubscript{2}Te\textsubscript{3}, the impact of including many-body heat current causes a decrement of 29\% in $\kappa_{L}$. The result suggests that the correction to the heat current formula only indicates the potential pitfall of the original heat current value and its auto-correlation, and does not guarantee the behavior of its auto-correlation function and its impact on $\kappa_{L}$.

\par Extending the concept into 2D material, the Green-Kubo \textit{EMD} result shown in Fig. \ref{fig:sub7} reveals an increment of the heat current auto-correlated thermal conductivity of graphene compared before and after the modification. The $\kappa_{L}$ rises 51\% from 1060 to 1605 W/mK which are considerably smaller than other reported \textit{MD} simulations by Fan \cite{fan2021}, Zhang \cite{zhang2011} and Gu \cite{Gu2011}, ranging from around 2300 to 2900 W/mK. The increment of $\kappa_{L}$ suggests that the integration of many-body contribution of heat flux plays a significant role into $\kappa_{L}$ of graphene.  The discrepancy in our \textit{EMD} result with other literature can be due to the size effect of our supercell. Similar \textit{MD} results were reported by Pereira and Donadio \cite{pereira2013} who found that graphene convergence behaves differently when compared to other materials due to failing in sampling low-frequency acoustic flexural mode at small simulation sizes.  A recent work by Fan et al. \cite{fan2017} further hints to the idea of a lack of flexural mode contribution as they decompose the contribution of in-plane and out-of-plane components to $\kappa_{L}$ of graphene. It is revealed that the graphene thermal conductivity is dominated by out-of-plane mode up to 60-70\%. Increment in $\kappa_{L}$ after the correction implies an underestimation of the in-plane $\kappa_{L}$ contributions in the original MTP formula. Despite a better agreement with previous theoretical prediction could be expected by a larger simulation supercell, the computation time is formidable with current computing resources. As we are focusing on the missing of accounting effect of the many-body heat current formula to the MTP potential in the existing interface packages, the current result is sufficient to demonstrate the presence of a many-body heat current contribution.

\par The final example is BAs which we also find an interesting trend for the change in $\kappa_{L}$ based on the Green-Kubo method. As shown in Fig. \ref{fig:sub8}, a decrease of 19\% from 1260 to 1017 W/mK of $\kappa_{L}$ is recorded. Compared to the experimental results of Tain et al. \cite{tain2018} who reported $\kappa_{L}$ of BAs at room temperature varies from 450$\pm$60 to 1160$\pm$130 W/mK across various measurement locations on multiple samples, our result based on MD simulation lies within the range of the measurements. A computation study based on perturbation theory and phonon Boltzmann's transport equation (PBTE) reported a $\kappa_{L}$ of around 1400 W/mK incorporating both three- and four-phonon interactions. The difference between Green-Kubo and \textit{PBTE} methods on $\kappa_{L}$ may be related to the higher-order phonon scattering processes unfolded in \textit{MD} simulations or phonon-boundary scatterings from the size effect. However, as shown in Fig. \ref{fig:sub4}, the difference of BAs overall cumulative heat current between the original and modified formula is trivial. This indicates that despite an insignificant change in terms of the averaged absolute value of the heat current between the original and the many-body heat current formula, the heat current auto-correlation function can still be altered and distinguishable between the two models.
\\
\section{\label{sec:level1_concl}Conclusion\protect}

\par In summary, we demonstrated the significance of many-body interaction when considering the heat current of a model system and MTP was selected as the workhorse of MD simulation. A modification of the LAMMPS/MLIP interface was implemented based on the generalized many-body heat current formula. Four examples namely PbTe, amorphous Sc\textsubscript{0.2}Sb\textsubscript{2}Te\textsubscript{3}, graphene, and BAs, covered an extensive range in terms of order of magnitude in $\kappa_{L}$ and crystal geometry complexity. All examples except BAs show a large difference between the MTP evaluated heat current and the actual overall heat current of the NEMD simulation. We further presented that the modification improves the agreement of the MTP calculated heat current and the overall heat current of the MD simulation, suggesting an underestimation of the heat current value without considering the many-body contribution.

\par The $\kappa_{L}$ comparison of the four examples reveals the significant importance of the corrections. Both low $\kappa_{L}$ materials (PbTe and amorphous Sc\textsubscript{0.2}Sb\textsubscript{2}Te\textsubscript{3}) demonstrated better results with alignment to published experimental and computational data, while graphene and BAs results lie within a reasonable range of values. Our calculation suggested that the rectified heat current can impact the computed $\kappa_{L}$ up to 64\%, especially for low $\kappa_{L}$ materials. This work illustrates the significance of adopting the generalized many-body heat current formula and its underlying influence on the thermal conductivity computed based on MD simulation approaches using the MTP heat current operator.


\begin{acknowledgments}
This work is supported by the Research Grants Council of Hong Kong (C7002-22Y, 17318122 and C6020-22GF). The authors are grateful for the research computing facilities offered by ITS, HKU.
\end{acknowledgments}

\nocite{*}

\bibliography{main}

\end{document}